\documentclass[12pt,a4,epsf]{article} \usepackage{cite}
\usepackage{graphicx} \usepackage{amssymb} \global\arraycolsep=2pt
\input{epsf} 
\usepackage{graphicx}
\def\@fmsl@sh#1#2#3{\m@th\ooalign{$\hfil#1\mkern#2/\hfil$\crcr$#1#3$}}
 \def\eq#1\en{\begin{equation}#1\end{equation}}
\def\s[#1,#2]{[#1\stackrel{\star}{,}#2]}
\def\sx[#1,#2]{[#1\stackrel{\star_{x}}{,}#2]} \def\pp#1{\partial_#1}

\def\lrDmu{\stackrel{\leftrightarrow}{D_\mu}}
\def\lrDnu{\stackrel{\leftrightarrow}{D^\nu}}
\begin{document}
\makeatletter
\def\fmslash{\@ifnextchar[{\fmsl@sh}{\fmsl@sh[0mu]}}
\def\fmsl@sh[#1]#2{%
  \mathchoice
    {\@fmsl@sh\displaystyle{#1}{#2}}%
    {\@fmsl@sh\textstyle{#1}{#2}}%
    {\@fmsl@sh\scriptstyle{#1}{#2}}%
    {\@fmsl@sh\scriptscriptstyle{#1}{#2}}}
\def\@fmsl@sh#1#2#3{\m@th\ooalign{$\hfil#1\mkern#2/\hfil$\crcr$#1#3$}}
\makeatother

\thispagestyle{empty}
\begin{titlepage}
\begin{flushright}
hep-ph/0401097 \\
CALT-68-2473\\
January 13, 2004
\end{flushright}

\vspace{0.3cm}
\boldmath
\begin{center}
  \Large {\bf What are the Bounds on Space-Time Noncommutativity?}
\end{center}
\unboldmath
\vspace{0.8cm}
\begin{center}
  {\large Xavier Calmet\footnote{
email:calmet@theory.caltech.edu}}\\ \end{center}
\begin{center}
{\sl California Institute of Technology, Pasadena, 
California 91125, USA}
\end{center}
\vspace{0.8cm}
\begin{abstract}
\noindent
In this article we consider the bounds on the noncommutative nature of
space-time. We argue that these bounds are extremely model
dependent. In the only phenomenologically viable framework, i.e. when
the fields are taken to be in the enveloping of the Lie algebra, the
constraints are fairly loose and only of the order of a few TeV.
\end{abstract}  

\end{titlepage}
The aim of this work is to discuss the bounds on the noncommutative
nature of space-time. We will argue that these bounds are extremely
model dependent and in particular depend largely on whether the
noncommutative fields are Lie algebra valued or in the enveloping
algebra. For reasons that will be explained later, the only
phenomenological viable approach is the one where fields are assumed
to be in the enveloping algebra. It turns out that in that case the
bounds are fairly loose and are of the order of a few TeV only.

The idea that space-time might be noncommutative at short distances is
not new and can traced back to Heisenberg \cite{Heisenberg}, Pauli
\cite{Pauli} and Snyder \cite{Snyder:1946qz}. This idea was taken very
seriously recently because noncommutative coordinates were found in a
specific limit of string theory. This is nevertheless not the only
motivation to study Yang-Mills theories on noncommutative spaces. In
the early days of quantum field theories, it was thought that a
fundamental cutoff might be useful to regularize the infinities
appearing in these theories. Nowadays it is understood that gauge
theories describing the strong and electroweak interactions are
renormalizable and thus infinities cancel, but it might still be
useful to have a fundamental cutoff to make sense of a quantum theory
of gravity, whatever this might be. A more pragmatic approach is that
space-time could simply be noncommutative at short distances in which
case one has to understand how the standard model can emerge as a low
energy model of a Yang-Mills theory formulated on a noncommutative
space-time.

The simplest noncommutative relations one can study are
\begin{equation} \label{relord}
  [\hat x^\mu,\hat x^\nu] \equiv \hat x^\mu \hat x^\nu- \hat x^\nu \hat x^\mu 
  =i \theta^{\mu \nu}, \qquad \theta^{\mu\nu}\in\mathbb{C}. \label{canonical}
\end{equation}
Postulating such relations implies that Lorentz covariance is
explicitly broken.  These relations also imply uncertainty relations
for space-time coordinates:
\begin{equation}
  \Delta x^\mu \Delta  x^\nu \ge \frac{1}{2} | \theta^{\mu \nu}|,
\end{equation}
which are a reminiscence of the famous Heisenberg uncertainty
relations for momentum and space coordinates.  Note that $\theta^{\mu
\nu}$ is a dimensional full quantity, dim($\theta^{\mu
\nu}$)$=$mass$^{-2}$. If this mass scale is large enough, $\theta^{\mu
\nu}$ can be used as an expansion parameter like $\hbar$ in quantum
mechanics. We adopt the usual convention: a variable or function with
a hat is a noncommutative one. It should be noted that the relations
(\ref{relord}) are very specific and other relations have been
considered. Other examples are Lie structures $[\hat x^\mu, \hat
x^\nu]=i f^{\mu \nu}_\alpha \hat x^\alpha$ and quantum plane
structures $[\hat x^\mu, \hat x^\nu]=i C^{\mu \nu}_{\alpha \beta} \hat
x^\alpha \hat x^\beta$. It is known how to formulate Yang-Mills
theories on a generic Poisson structure \cite{Calmet:2003jv}.

The aim of this work is to discuss the bounds on space-time
noncommutativity appearing in the literature. It should be noted that
most bounds on the noncommutative nature of space-time come from
constraints on Lorentz invariance. These constraints are extremely
model dependent. There are different approaches to gauge field theory
on noncommutative spaces. The first approach is motivated by string
theory, see e.g \cite{Douglas:2001ba} for a review. It is
non-perturbative in $\theta$ and the non-local property of the
interactions is manifest. Fields are taken as usual to be Lie algebra
valued. Unfortunately it turns out that this approach suffers from a
number of drawbacks that make it unsuitable to build realistic models
for the electroweak and strong interactions.

If fields are assumed to be Lie algebra valued, it turns out that only
U(N) structure groups are conceivable because the commutator
\begin{eqnarray}
\s[\hat\Lambda,\hat\Lambda'] = \frac{1}{2}\{\hat
\Lambda_a(x)\stackrel{\star}{,} \hat \Lambda'_b(x)\}[T^a,T^b] +
\frac{1}{2}\s[\hat \Lambda_a(x),\hat \Lambda'_b(x)]\{T^a,T^b\}
\label{com}
\end{eqnarray}
of two Lie algebra valued noncommutative gauge parameters
$\hat\Lambda = \Lambda_a(x) T^a$ and $\hat\Lambda' = \Lambda'_a(x)
T^a$ only closes in the Lie algebra if the gauge group under
consideration is U(N) and if the gauge transformations are in the
fundamental representation of this group. But, this approach cannot be
used to describe particle physics since we know that SU(N) groups are
required to describe the weak and strong interactions. Or at least
there is no obvious way known to date to derive the standard model as a
low energy effective action coming from a U(N) group. 

Furthermore it turns out that even in the U(1) case, charges are
quantized \cite{Hayakawa:1999yt,Hayakawa:1999zf} and it thus
impossible to describe quarks. This problem is obvious if one writes
the field strength tensor explicitly:
\begin{eqnarray}
F_{\mu \nu}=i\s[D_\mu,D_\nu]=\partial_\mu A_\nu -
\partial_\nu A_\mu-i \s[A_\mu,A_\nu].
\end{eqnarray}
The commutator $\s[A_\mu,A_\nu]$ does not vanish even for a U(1)
gauge group, the choice of charges introduced in the theory is very
limited namely $\pm$1 or 0.

There is a framework that enables to address these problems
\cite{Madore:2000en,Jurco:2000ja,Jurco:2001rq,Calmet:2001na}. The
aim of this new approach is to derive low energy effective actions for
the noncommutative theory which is too complicated to handle. The
matching of the noncommutative action to the low energy action on a
commutative space-time is done in two steps. First the noncommutative
coordinates are mapped to usual coordinates, the price to pay is the
introduction of a star product \cite{Moyal:sk}.  Secondly the
noncommutative fields are mapped to commutative fields by means of the
Seiberg-Witten maps. The Seiberg-Witten maps \cite{Seiberg:1999vs}
have the remarkable property that ordinary gauge transformations
$\delta A_\mu = \pp\mu \Lambda + i[\Lambda,A_\mu]$ and $\delta \Psi =
i \Lambda\cdot \Psi$ induce noncommutative gauge transformations of
the fields $\hat A$, $\hat \Psi$:
\begin{eqnarray}
\delta \hat A_\mu = \hat\delta \hat A_\mu,
\qquad \delta \hat \Psi = \hat\delta \hat \Psi. \label{gequiv}
\end{eqnarray}

The low energy action is local in the sense that there is no UV/IR
mixing in that approach. The noncommutative nature of space-time is
encoded in the higher order operators that enter the theory. The basic
assumption is that the noncommutative fields are not Lie algebra
valued but are in the enveloping of the algebra:
\begin{eqnarray} \label{envalg}
\hat\Lambda = \Lambda^0_a(x) T^a +  \Lambda^1_{ab}(x) :T^a T^b:
+ \Lambda^2_{abc}(x) :T^a T^b T^c: + \ldots
\end{eqnarray}
where $:\;:$ denotes some appropriate ordering of the Lie algebra
generators. One can choose, for example, a symmetrically ordered basis
of the enveloping algebra, one then has $:T^a:=T^a$ and $:T^a T^b:=
\frac{1}{2} \{T^a, T^b \}$ and so on. Taking fields in the enveloping
of the algebra allows to consider SU(N) groups since in that case the
relation (\ref{com}) can close. At first sight it seems that one has
introduced a infinity number of degrees of freedom. It turns out that
all fields appear in (\ref{envalg}) can be expressed in terms of the
classical gauge parameter. Higher order term in (\ref{envalg}) are
assumed to be suppressed by higher powers of $\theta$.

Expanding to linear order in $\theta$ the star product and the
noncommutative fields, one obtains the action:
\begin{eqnarray}
\label{actionqed}
\int  \bar{\hat\Psi}\star (i \gamma^\mu { \hat D}_\mu-m) \hat \Psi
 d^4x&=& \int
 \bar{\psi} (i \gamma^\mu D_\mu- m)\psi d^4x \\ \nonumber
&&-\frac{1}{4} \int 
\theta^{\mu \nu}
 \bar{\psi} F_{\mu \nu} (i \gamma^\alpha D_\alpha -m )\psi d^4x 
\\ && \nonumber
-\frac{1}{2}  \int \theta^{\mu \nu}
 \bar{\psi} \gamma^\rho F_{\rho \mu} i D_\nu \psi  d^4x 
\\ \nonumber 
-\frac{1}{4} \hat F_{\mu \nu } \star
 \hat F^{\mu \nu} d^4 x&=&-\frac{1}{4} \int  F_{\mu \nu }
 F^{\mu \nu } d^4x
\\ \nonumber && +\frac{1}{8} 
\int \theta^{\sigma \rho } F_{\sigma \rho }F_{\mu \nu } F^{\mu \nu } d^4x \\ \nonumber
&&-\frac{1}{2}\int  \theta^{\sigma \rho } F_{\mu \sigma}F_{\nu \rho} F^{\mu \nu} d^4x.
\end{eqnarray} 

There are a number of difficulties which have to be addressed in order
to formulate the standard model on a noncommutative space-time.  These
problems have been solved in \cite{Calmet:2001na}.

The first problem is that one cannot introduce three different
noncommutative gauge potentials. The reason is that noncommutative
gauge invariance is linked to the invariance of the covariant
coordinates $\hat X^\mu = \hat x^\mu+\hat B^\mu$. The Yang-Mills
potential $A_\mu$ is related to $B^\mu$ by $B^\mu=\theta^{\mu \nu}
A_\nu$, i.e. gauge transformations are related to transformations of
the covariant coordinate. The solution is to introduce a master field:
$V_\mu=g' A_\mu+g B_\mu+g_S G_\mu$ that contains all the gauge
potential of the structure group SU(3)$\times$SU(2)$\times$U(1) and to
performed a Seiberg-Witten map for
$\hat V_\mu$. Note that a generalized gauge transformation is also
introduced $\Lambda=g'\alpha(x) Y + g \alpha_L(x)+g_s \alpha_s(x)$, with the Seiberg-Witten map $\hat \Lambda = \Lambda +\frac{1}{4}  \theta^{\mu \nu}
\{V_\nu,\partial_\mu \Lambda\}+{\cal O}(\theta^2)$.

The approach presented in \cite{Calmet:2001na} offers a very natural
problem to the charge quantization problem. One introduces $n$
different noncommutative hyperphotons, one for each charge entering
the model:
\begin{eqnarray}
\hat \delta \hat a_i^{(n)} =\partial_i \hat \lambda^{(n)} + 
i [\hat \lambda^{(n)},\hat a_i^{(n)}]
\end{eqnarray}
with
\begin{eqnarray}
\hat \delta \hat{\Psi}^{(n)}= i e q^{(n)} \hat \lambda^{(n)} \star \hat \Psi^{(n)}.
\end{eqnarray}
At first sight, it seems that this implies the existence of $n$
photons in nature, i.e. that the theory has too many degrees of
freedom, but once again the Seiberg-Witten maps can be used to reduce
the amount of degrees of freedom. It turns out that these $n$
noncommutative hyperphotons have the same classical limit $a_i$:
\begin{eqnarray}
\hat a_i^{(n)}=a_i- e q^{(n)} \frac{1}{4}\theta^{kl}\{a_k,\partial_l a_i + f_{li}\}+{\cal O}(\theta^2)
\end{eqnarray}
i.e. there is only one classical photon.

Another problem are the Yukawa couplings: a noncommutative field can
transform on the left-hand side or on the right-hand side and this
makes a difference. This is an obvious complication for Yukawa
couplings. For example $\bar{\hat L} \star \hat \Phi \star \hat e_R$
is not invariant under a noncommutative gauge transformation if $\hat
\Phi$ transforms only on the right-hand side or only on the left-hand
side. The solution is to assume that it transforms on both sides to
cancel the transformations of the SU(2) doublet and of the SU(2)
singlet fields, $\bar{\hat L} \star \rho_L(\hat \Phi) \star \hat e_R$
with $\rho_L(\hat \Phi)=\Phi[\phi,-\frac{1}{2}g'{\cal A}_\mu+g
B_\mu,g'{\cal A}_\mu]$ and $\widehat\Phi[\Phi,A,A'] = \Phi +
\frac{1}{2}\theta^{\mu\nu} A_\nu \Big(\partial_\mu\Phi -\frac{i}{2}
(A_\mu \Phi - \Phi A'_\mu)\Big)+\frac{1}{2}\theta^{\mu\nu}
\Big(\partial_\mu\Phi -\frac{i}{2} (A_\mu \Phi - \Phi
A'_\mu)\Big)A'_\nu$. Note that it is possible to couple neutral
particles to the photon in a gauge invariant way.

It should be noted that the form of the operators that enter the
effective theory is very severely constrained by the noncommutative
gauge invariance. Naively one could guess that an operator $m
\theta^{\mu\nu} \bar \Psi \sigma_{\mu\nu} \Psi$ could appear in the
low energy effective action \cite{Anisimov:2001zc}. After all the
Wilsonian approach to effective theories teaches us that an operator
not forbidden by a symmetry will enter the theory with potentially a
coefficient of order one. But, it is absolutely not clear that such an
operator is compatible with the noncommutative gauge invariance, and
might thus be simply forbidden. One might argue that it is generated
by a term $m \theta^{\mu\nu} \bar{\hat \Psi} \sigma_{\mu\nu}\star \hat
\Psi$ that is invariant under noncommutative gauge transformations,
but such an operator makes little sense since $\theta^{\mu\nu}$ only
enters the theory through the star product and the Seiberg-Witten maps
of the fields. One would have to show that such an operator can be
generated at the loop level on the noncommutative side, which seems
doubtful since the noncommutative action is non-perturbative in
$\theta$.  One has to be very careful when effective theory arguments
are applied to these models since it is very difficult to keep track
of the fundamental symmetry which is the noncommutative gauge
invariance.

Another source of model dependence originates in the choice of the
definition of the trace in the enveloping algebra and of the
representation of the noncommutative field strength $\widehat
F^{\mu\nu}$. The action for non-Abelian noncommutative gauge bosons is
$$ S_\mathrm{gauge} = -\frac{1}{2}\int d^4 x \, \mbox{\bf Tr}
\frac{1}{\mbox{\bf G}^2} \widehat F_{\mu\nu} \star \widehat
F^{\mu\nu}, $$ with the noncommutative field strength $\widehat
F_{\mu\nu}$, an appropriate trace $\mbox{\bf Tr}$ and an
operator~$\mbox{\bf G}$. This operator must commute with all
generators ($Y$, $T_L^a$, $T_S^b$) of the gauge group so that it does
not spoil the trace property of $\mbox{\bf Tr}$.

The operator $\mbox{\bf G}$ is in general a function of $Y$ and the
Casimir operators of SU(2) and SU(3). However, due to the
assignment of hypercharges in the standard model it is possible to
express~$\mbox{\bf G}$ using $Y$ and six constants $g_1$, \ldots,
$g_6$ corresponding to the six multiplets.  In the classical limit
only certain combinations of these six constants, corresponding to the
usual coupling constants $g'$, $g$ and $g_S$ are relevant. The
relation is given by the following equations: 
\begin{eqnarray}
\frac{1}{g_1^2} + \frac{1}{2 g_2^2} + \frac{4}{3 g_3^2} + \frac{1}{3
g_4^2} + \frac{1}{6 g_5^2} + \frac{1}{2 g_6^2} = \frac{1}{2g'^2}\\
\nonumber \frac{1}{g_2^2} + \frac{3}{g_5^2} + \frac{1}{g_6^2} =
\frac{1}{g^2} \\ \nonumber \frac{1}{g_3^2} + \frac{1}{g_4^2}
+\frac{2}{g_5^2} = \frac{1}{g_S^2}. \label{simplex}
\end{eqnarray}

These three equations define for fixed $g'$, $g$ and $g_S$ a
three-dimensional simplex in the six-dimensional moduli space spanned
by $1/g_1^2$, \ldots, $1/g_6^2$. The remaining three degrees of
freedom become relevant at order $\theta$ in the expansion of the
noncommutative action. The traces corresponding to triple gauge boson
vertices: \begin{eqnarray} \mbox{\bf Tr} \frac{1}{\mbox{\bf G}^2} Y^3
= -\frac{1}{g_1^2} - \frac{1}{4 g_2^2} + \frac{8}{9 g_3^2} -
\frac{1}{9 g_4^2} + \frac{1}{36 g_5^2} + \frac{1}{4 g_6^2}
\end{eqnarray}
\begin{eqnarray}
 \mbox{\bf Tr} \frac{1}{\mbox{\bf G}^2} Y
T_L^a T_L^b = \frac{1}{2}\delta^{ab}\left(-\frac{1}{2g_2^2} +
\frac{1}{2g_5^2} + \frac{1}{2g_6^2} \right) 
\end{eqnarray}

\begin{eqnarray}  \mbox{\bf Tr}
\frac{1}{\mbox{\bf G}^2} Y T_S^c T_S^d = \frac{1}{2}\delta^{cd}\left(
\frac{2}{3g_3^2} - \frac{1}{3g_4^2} +\frac{1}{3g_5^2} \right) 
\end{eqnarray}
are of particular interest.
One consequence is that the triple photon vertex cannot be used to
bound space-time noncommutativity. While such an interaction can be
seen as a smoking gun of space-time noncommutativity, the bounds
obtained are model dependent and only constrain a combination of
$\theta^{\mu\nu}$ and of an unknown coupling constant. It is worth
noting that most collider studies have considered modifications of the
gauge sector to search for space-time noncommutativity see
e.g. \cite{Rizzo:2002yr,Behr:2002wx}. While these channels and rare
decays are interesting from the discovery point of view, they cannot
be used to bound the noncommutative nature of space-time itself.

The only model independent part of the effective action is the fermionic
sector. There are two types of bounds in the literature that are
relevant to the case where fields are taken to be in the enveloping
algebra.

The first relevant study is that of Carroll et
al. \cite{Carroll:2001ws}.  They replace $F_{\mu\nu} \to f_{\mu\nu} +
F_{\mu\nu}$ in eq. (\ref{actionqed}), where $f_{\mu\nu}$ is understood
to be a constant background field and $F_{\mu\nu}$ now denotes a small
dynamical fluctuation.

Keeping only terms up to quadratic order in the fluctuations
and performing a physically irrelevant rescaling
of the fields $\Psi$ and $A_\mu$
to maintain conventionally normalized kinetic term, they obtained

\begin{eqnarray}
L &=& 
\frac{1}{2} i \bar{\Psi} \gamma^\mu \lrDmu \Psi
- m\bar{\Psi} \Psi
- \frac{1}{4} F_{\mu\nu} F^{\mu\nu}
\nonumber\\
&&
+ \frac{1}{2}i c_{\mu\nu} \bar{\Psi} \gamma^\mu \lrDnu \Psi
- \frac{1}{4} {k_F}_{\alpha\beta\gamma\delta} F^{\alpha\beta} 
F^{\gamma\delta}.
\label{fqed}
\end{eqnarray}
They have replaced, in this equation, the charge $q$ in the covariant
derivative with a scaled effective value
\begin{eqnarray}
q_{\rm eff} = (1 + \frac{1}{4} q f^{\mu\nu}\theta_{\mu\nu}) q.
\label{charge}
\end{eqnarray}
The coefficients $c_{\mu\nu}$ and 
${k_F}_{\alpha\beta\gamma\delta}$ are
\begin{eqnarray}
c_{\mu\nu} &=& - \frac{1}{2} q f_\mu^{\lambda} \theta_{\lambda \nu},
\nonumber\\
{k_F}_{\alpha\beta\gamma\delta} &=& 
- q f_\alpha^{\lambda} \theta_{\lambda\gamma}\eta_{\beta\delta}
+ \frac{1}{2} q f_{\alpha\gamma} \theta_{\beta\delta} 
- \frac{1}{4} q f_{\alpha\beta} \theta_{\gamma\delta}
\nonumber\\
&&
- (\alpha \leftrightarrow \beta)
- (\gamma \leftrightarrow \delta)
+ (\alpha\beta \leftrightarrow \gamma\delta).
\label{coeffs}
\end{eqnarray}
${k_F}_{\alpha\beta\gamma\delta}$ is only very weakly constrained by
experiments.  That constraint would be model dependent since these
coefficient depends on the choice of the representation for the
noncommutative gauge fields and thus on the way the trace in the
enveloping algebra is defined. On the other hand the coefficient
$c_{\mu\nu}$ is accessible through clock comparison studies and is
directly related to the fermionic sector of the action.  Carroll et
al. obtain the bounds $|\theta^{YZ}|,|\theta^{ZX}| \le (10 \
\mbox{TeV})^2$ using a rather crude model for the $^9$Be nucleus wave
function.

The other constraint on space-time noncommutativity relevant to the
case where the noncommutative fields are taken to be in the enveloping
algebra comes from a study by Carlson at
al. \cite{Carlson:2001sw}. They study noncommutative QCD at the one
loop order. They considered the one loop correction to the quark mass
and wavefunction renormalization and performed their calculation using
the low energy effective action (\ref{actionqed}). The one loop
expression needs to be regulate, the authors of \cite{Carlson:2001sw}
choose to do so by a Pauli-Villars regularization procedure. While
they are very careful not to break the classical gauge invariance,
there is a priori no guaranty that such a procedure respects the
noncommutative gauge invariance. But, let us assume that the
Pauli-Villars regulator respects both symmetries. The result obtain in
\cite{Carlson:2001sw} is, keeping just the $O(\theta)$ terms,
\begin{equation}
i {\cal M}(\lambda^2,M^2) = 
               {2\over 3} g^2 \{ (\not\! p -m), \sigma_{\mu\alpha} \}       
\times \int {(dq)\over (q^2- \lambda^2) 
                         \left( (p+q)^2-M^2 \right)   }
            q^\alpha \theta^{\mu\nu} (p+q)_\nu   \ .
\end{equation}
The Pauli-Villars regulated amplitude then given by 
${\cal M} \rightarrow {\cal M}(0,m^2) - {\cal M}(\Lambda^2,m^2) - 
{\cal M}(0,\Lambda^2) + {\cal M}(\Lambda^2,\Lambda^2)$, where $\Lambda$
is a large mass scale.  Their result is
\begin{equation}                             \label{result}
{\cal M} = {g^2 \over 96 \pi^2} \bigg(
       \big\{ (m - \not\! p),  \ 
       \Lambda^2 \theta^{\mu\nu} \sigma_{\mu\nu}   \big\} 
-{2\over 3} \big\{ (m - \not\! p),  \ 
       p_\mu \theta^{\mu\nu} \sigma_{\nu\tau} p^\tau \ln{\Lambda^2}  
                         \big\}  \bigg) \ , 
\end{equation}
\noindent for the term leading in $\Lambda$ for each Dirac structure.
The authors of \cite{Carlson:2001sw} considered the three operators 
\begin{equation} \label{ops}
m \theta^{\mu\nu} \bar q \sigma_{\mu\nu} q , \,\,\,\,\,  
\theta^{\mu\nu} \bar q \sigma_{\mu\nu} \not\!\! D q , \,\,\,\,\,
\mbox{ and } \,\,\,\,\,
\theta^{\mu\nu} D_\mu \bar q \ \sigma_{\nu\rho} D^\rho q,  
\end{equation}
and obtained, using the first of these operators, the bound:
\begin{equation}\label{eq:bound}
\theta \Lambda^2 \mbox{\raisebox{-1.0ex} 
{$\stackrel{\textstyle ~<~} {\textstyle \sim}$}} 10^{-29} \,\,\, ,
\end{equation}
where $\Lambda$ is an ultraviolet regularization scale.  But, these
operators enter the game in a very specific combination. A closer look
at (\ref{result}) reveals that the matrix element is vanishing. Since
we are working just to first order in the operators (\ref{ops}) the
QCD equations of motion $(i \fmslash{D} -m)q=0$ can be used
\cite{Politzer:1980me}. This invalidates the bound (\ref{eq:bound})
and is a very strong indication these operators are forbidden by the
noncommutative gauge invariance.

The bounds on the noncommutativity of space-time are thus fairly loose
if fields are taken to be in the enveloping algebra, and of the order
of 10 TeV. Much more effort has to be invested to derive bounds on the
noncommutative nature of space-time. It is important to realize that
any bound is framework dependent and even in a given framework there
is, most of the time, some model dependence. We have a clear idea of
what signal would have to be interpreted as an evidence for the
noncommutativity of space-time, on the other hand bounding the
noncommutative parameter $\theta^{\mu\nu}$ is a very difficult task.

The fact that the bounds are on the order of 10 TeV should not be
taken as an indication that colliders studies are useless. It is
conceivable that $\theta^{\mu\nu}$ is not a constant but a more
complicated function. As it has been argued in \cite{Calmet:2003jv},
the higher order operators that describe the noncommutative nature of
space-time might very well be energy-momentum dependent and thus only
become relevant at high energies or equivalently at short
distance. This should be a very strong motivation to study more model
independent contributions to particle reactions that can be studied at
the next generation of colliders. Some work in that direction
\cite{Iltan:2002fa,Mahajan:2003ze,Iltan:2002ip} has already been done,
but much more remains to be done.

\section*{Acknowledgments}
The author is grateful to H.~D.~Politzer, M.~Ramsey-Musolf and
M.~B.~Wise for enlightening discussions. Insightful discussions with
C. Carone and R. Lebed about their work are also gratefully
acknowledged.

\end{document}